# Efficiency limit of intermediate band $Al_xGa_{1-x}As$ solar cell based on $Al_yGa_{1-y}Sb$ type-II quantum dots embedded outside of the depletion region

A. Kechiantz[1,2,*], A. Afanasev[2,**], and J.-L. Lazzari[3,***]

[1]Institute of Radiophysics and Electronics, National Academy of Sciences, Ashtarak, ARMENIA
[2]Department of Physics, The George Washington University, Washington, DC, USA
[3]Centre Interdisciplinaire de Nanoscience de Marseille, CNRS – Aix-Marseille Université, FRANCE

## 1) Context / Study motivation

Highly efficient harvesting of sunlight is a pressing issue for energy supply to remote limited-size devices [1]. Techniques that are generally accepted in photovoltaic industry use multiple-junction solar cells with photoactive p-n-junctions connected in series by tunnel-junctions for adding generated photovoltages. However, recombination at tunnel-junctions reduces the efficiency, for instance, to 2/3 of the detailed balance limit even in the best triple-junction solar cells.

The intermediate band (IB) solar cell is another concept of highly efficient cells. While the multi-junction solar cells exploit absorption of concentrated low-energy photons for generation of additional photovoltage in series connected p-n-junctions, the IB concept uses nonlinear effect of two photon absorption essentially enforced with concentration of such photons for generation of additional photocurrent in single p-n-junction cells [2]. Conversion efficiency of such IB GaAs cells may achieve 60% [3].

There are two opportunities for IB location in solar cells, either within [3] or outside the depletion regions [2]. The concept does not justify location of IB states in the cell. However, recent experimental study of InAs quantum dots (QDs) as IB sandwiched between n-doped and p-doped GaAs layers has shown that such IB cells generate smaller photovoltage than the reference GaAs cells because of additional dark current generated in InAs QDs [3]. Also it became clear that InAs is not the best counterpart of GaAs in IB solar cells [2].

In this work we report on IB potential of self-organized strained type-II $Al_yGa_{1-y}Sb$ QDs for operation in GaAs solar cells. To demonstrate our results, we focus on QDs embedded outside the build-in field of the p-n-junction and apply electronic properties known from capacitance–voltage and deep level spectroscopy study of self-organized GaSb QDs embedded in GaAs [4].

## 2) Description of approach and techniques

Figure 1a displays the structure of novel IB GaAs solar cell formed by a graded epitaxial layer of alternating $Al_yGa_{1-y}Sb$ QDs and p-doped $Al_xGa_{1-x}As$ spacers grown over a thin p-doped GaAs buffer layer covering n-doped GaAs substrate. Here $x$ rises from 0 at the buffer to about 0.3. The epitaxial layer is as thin as the electron diffusion length. Such band gap variation (from $1.43 eV$ to $1.83 eV$) creates about $1 kV/cm$ pulling-field in the grade layer that accelerates electrons up to about $10^6 cm/s$ velocity towards the p-n-junction. For $1 ns$ of inter-band recombination time, electrons can pass through $10\mu$-thick grade layer and reach the p-n-junction. Since the intra-band relaxation time is 50 $ps$ [5], also carriers lifted into continuous conduction and valence bands of $10 nm$-thick QDs swiftly escape and relax within $Al_xGa_{1-x}As$ spacer.

Figure 1b illustrates energy band diagram of the IB solar cell. The orange dot-line is the edge of continuous energy spectra in the valence band of type-II QDs. Black layers are ohmic contacts. The red and blue arrows indicate generation-recombination transitions. While $j_{CV}$ transitions generate the conventional photocurrent, $j_{IV}$ and $j_{IC}$ transitions balance each other, $j_{IV} = j_{IC}$, and generate the additional photocurrent of the IB cells. To be in balance, such electron transitions charge QDs positive so that a shell barrier $\varepsilon_L$ rises around QDs. Since the shell barrier spatially separates confined electron-hole pairs, it also suppresses recombination in QDs by blocking hole-transfer from the spacer into QDs. We have used the microsecond lifetime for non-radiative recombination of such electrons-hole pairs spatially separated in the vicinity of the shell barrier.

Figure 2 displays evaluated conversion efficiency of the reference and IB solar cells, red (b and d) and blue (a and c) curves, calculated for one and 500 sun illumination, dot (a and b) and solid (c and d) curves, respectively.

## 3) Results / Conclusions / Perspectives

Our study has shown that concentration of incoming photons is an important factor for achieving high conversion efficiency. Concentration first modifies the charge confined in QDs and the shell barrier. Then it stabilizes the shell barrier. Under about 500 sun concentration holes fill about the half of ground QD confined states and get into a condition that generation essentially overcomes recombination through QDs. Such transformation of QDs raises the efficiency of IB solar cell by 14% from 28% to 42% as shown in Figure 2. The same concentration improves efficiency of the reference cell by 5% only, from 29% to 34%.

Our study has shown that QDs assist in generation of the additional photocurrent. However, like artificial atoms QDs may easily convert their ground confined state into fast recombination level that promptly achieves equilibrium with conduction or valence bands. Such level also decreases the conventional photocurrent of IB solar cells.

In conclusion the important advantage of proposed IB solar cell design is that the "outside IB" does not assist generation of additional leakage current through QDs. Carriers confined in such QDs are far from the built-in electric field of the depletion layer and, hence, cannot escape from QDs into continuous bands without solar or thermal photon assistance. Recombination through QDs is a major factor that limits efficiency of solar cells based on QDs buried within the depletion region. Our proposal will help to solve this problem.

**ACKOWLEDGMENT:**

A. Kechiantz and J.-L. Lazzari carried out this work in the framework of the bilateral Armenian / French project, SCS/CNRS contract # IE-013 / # 23545. A. Kechiantz and A. Afanasev acknowledge support from The George Washington University.

\* E-mail (Kechiantz): kechiantz@gwu.edu
\*\* E-mail (Afanasev): afanas@gwu.edu
\*\*\* E-mail (Lazzari): lazzari@cinam.univ-mrs.fr

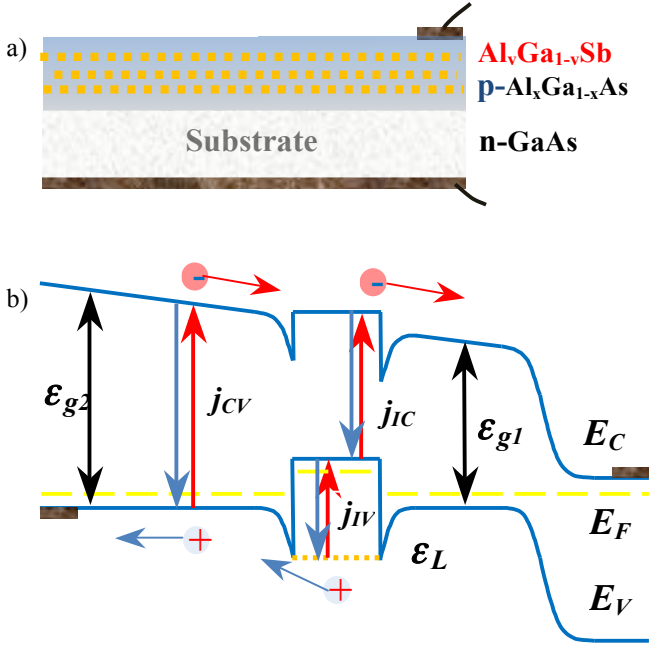

*Figure 1. Operation principle of the IB solar cell based on $Al_yGa_{1-y}Sb$ type-II QDs embedded outside of p-n-junction in graded $Al_xGa_{1-x}As$: a) structure; b) energy band diagram. The energy band offsets at QDs are 0.45eV in the valence and 0.1eV in the conduction bands [4].*

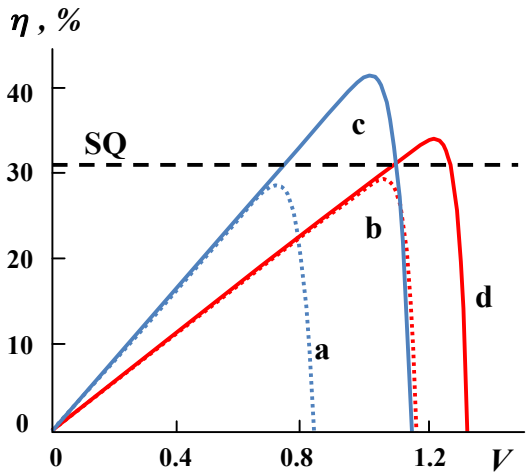

*Figure 2. Conversion efficiency of the reference (red curves) and IB (blue curves) solar cells for one sun (dashed curves) and 500 sun (solid curves) illumination. The dashed black line (SQ) is the Shockley-Queisser limit.*